\documentclass[journal]{IEEEtran}
\IEEEoverridecommandlockouts

\usepackage{amsmath}
\usepackage{amssymb}
\usepackage{amsfonts}
\usepackage{graphicx}
\usepackage{epsfig}
\usepackage{cite}
\usepackage{subfigure}
\usepackage{xcolor}
\usepackage[ruled,vlined,lined,ruled,linesnumbered]{algorithm2e}
\usepackage[font=small,skip=0pt]{caption}
\usepackage[belowskip=-10pt,aboveskip=0pt]{caption}
\begin{document}
\title{{\fontsize{23pt}{1em}\selectfont Distributed Charging Control in Broadband Wireless Power Transfer Networks}}
\vspace{-5ex}
\author{Suzhi~Bi, ~\IEEEmembership{Member,~IEEE} and Rui Zhang, ~\IEEEmembership{Senior Member,~IEEE}\\
                \thanks{This work was presented in part at the IEEE International Conference on Communications (ICC), London, June 8-12, 2015.}
                \thanks{The work of S. Bi is supported in part by the National Natural Science Foundation of China (project no. 61501303) and the Foundation of Shenzhen City (project no. JCYJ20160307153818306).}
        \thanks{S.~Bi is with the College of Information Engineering, Shenzhen University, Shenzhen, Guangdong, China 518060 (e-mail: bsz@szu.edu.cn).}
        \thanks{R.~Zhang is with the Department of Electrical and Computer Engineering, National University of Singapore, Singapore 117583 (e-mail:elezhang@nus.edu.sg). He is also with the Institute for Infocomm Research, A$^*$STAR, Singapore 138632.} }

\maketitle

\begin{abstract}
Wireless power transfer (WPT) technology provides a cost-effective solution to achieve sustainable energy supply in wireless networks, where WPT-enabled energy nodes (ENs) can charge wireless devices (WDs) remotely without interruption to the use. However, in a heterogeneous WPT network with distributed ENs and WDs, some WDs may quickly deplete their batteries due to the lack of timely wireless power supply by the ENs, thus resulting in short network operating lifetime. In this paper, we exploit frequency diversity in a broadband WPT network and study the distributed charging control by ENs to maximize network lifetime. In particular, we propose a practical voting-based distributed charging control framework where each WD simply estimates the broadband channel, casts its vote(s) for some strong sub-channel(s) and sends to the ENs along with its battery state information, based on which the ENs independently allocate their transmit power over the sub-channels without the need of centralized control. Under this framework, we aim to design lifetime-maximizing power allocation and efficient voting-based feedback methods. Towards this end, we first derive the general expression of the expected lifetime of a WPT network and draw the general design principles for lifetime-maximizing charging control. Based on the analysis, we then propose a distributed charging control protocol with voting-based feedback, where the power allocated to sub-channels at each EN is a function of the weighted sum vote received from all WDs. Besides, the number of votes cast by a WD and the weight of each vote are related to its current battery state. Simulation results show that the proposed distributed charging control protocol could significantly increase the network lifetime under stringent transmit power constraint in a broadband WPT network. Reciprocally, it also consumes lower transmit power to achieve nearly-perpetual network operation.
\end{abstract}


\section{Introduction}
The limited battery capacity is a major hurdle to the development of modern wireless technology. Frequent device battery outage not only disrupts the normal operation of individual wireless devices (WDs), but also significantly degrades the overall network performance, e.g., the sensing accuracy of a wireless sensor network. Conventional wireless systems require frequent recharging/replacement of the depleted batteries manually, which is costly and inconvenient especially for networks consisting of a large number of battery-powered WDs or operating under some special application scenarios, e.g., sensors embedded in building structure. Given stringent battery capacity constraints, minimizing energy consumption to prolong the WD operating lifetime is one critical design objective in battery-powered wireless systems. Using wireless communication networks for example, various energy-conservation schemes have been proposed, e.g., via transmit power management, energy-aware medium access control and routing selection, and device clustering, etc \cite{2002:Biyikoglu,2004:Younis,2007:Chen}.

The recent advance of wireless power transfer (WPT) technology provides an attractive alternative solution to power WDs over the air \cite{2015:Bi,2015:Lu,2016:Bi,2015:Ulukus,2014:Krikidis,2013:Zhou}, where WDs can harvest energy remotely from the radio frequency (RF) signals radiated by the dedicated energy nodes (ENs). Currently, with a transmit power of $3$ watts, tens of microwatts ($\mu$W) RF power can be transferred to a distance of several meters,\footnote{Based on the product specifications on the website of Powercast Co. (http://www.powercastco.com), with TX91501-3W power transmitter and P2110 Powerharvester receiver, the harvest RF power at a distance of $10$ meters is about $40\ \mu$W.} which is sufficient to power the activities of many low-power devices, such as sensors and RF identification (RFID) tags. Besides, WPT is fully controllable in its transmit power, waveforms, and occupied time/frequency resource blocks, thus can be easily adjusted in real-time to meet the energy demand of WDs. Its application can significantly improve the system performance and reduce the operating cost of a battery-powered wireless network. Due to the short operating range of WPT, a WPT network often needs to deploy \emph{multiple} ENs that are distributed in a target area to reduce the power transfer distance to the WDs within. Meanwhile, for radiation safety concern, densely deployed ENs are also necessary to reduce the individual transmit power of each EN for satisfying the equivalent isotropically radiated power (EIRP) requirement enforced by spectrum regulating authorities \cite{2015:Bi}. In light of this, we study in this paper the charging control for multiple distributed ENs in WPT networks.

The application of WPT also brings in a fundamental shift of design principle in energy-constrained wireless systems. Instead of being utterly energy-conservative in battery-powered systems, one can now prolong the device lifetime and meanwhile optimize the system performance by balancing the energy harvested and consumed. For point-to-point energy transfer, many techniques have been proposed to enhance the efficiency of WPT through, e.g., multi-antenna beamforming technique, WPT-tailored channel training/feedback, and energy transmitting/receiving antenna and circuit designs \cite{2013:Zhang,2015:Zeng,2015:Zeng1,2014:Xu}. From a network-level perspective, efficient methods have also been proposed to optimize both the long-term network placement (see e.g., \cite{2014:Huang1,2015:Bi1}) and real-time wireless resource allocation (see e.g., \cite{2014:Ju1,2014:Liu2,2015:Chen,2014:Zhou1,2016:Zhou,2013:Huang,2013:Nintanavongsa}) in WPT networks for optimizing the communication performance. Among them, one effective method is to exploit the \emph{frequency diversity} of multi-path fading channels in a broadband network \cite{2014:Zhou1,2013:Huang,2016:Zhou,2013:Nintanavongsa}. This is achievable by transmitting multiple energy signals on parallel frequency sub-channels that are separated at least by the channel coherence bandwidth. Intuitively, one can maximize the energy transfer efficiency in a point-to-point frequency-selective channel by allocating all transmit power to the strongest sub-channel. However, in the general case with multiple ENs and WDs with different sub-channel gains between each pair of EN and WD, there is a trade-off between ENs' energy efficiency and WDs' power balance in the transmit power allocation over frequency.

In this paper, we aim to optimize the transmit power allocation over frequency and time in a multi-EN and multi-WD broadband WPT network to maximize the network operating lifetime, which is a key performance metric of energy-constrained networks defined as the duration until a fixed number of WDs plunge into energy outage. A closely related topic is the design of lifetime-maximizing user scheduling in conventional battery-powered communication networks \cite{2004:Younis,2007:Chen} in the sense that the user scheduling determines the user priority to consume energy (transmit data), while the charging control problem considered in this paper determines the user priority to harvest more energy. Nonetheless, their designs differ significantly for two main reasons. On one hand, WPT to a particular WD will not cause detrimental co-channel interference to the others as in wireless information transmission (WIT), but can instead be exploited to boost the energy harvesting performance of all WDs \cite{2013:Liu}. On the other hand, the optimal power allocation to optimize the performance of WPT and WIT is fundamentally different. Using a point-to-point frequency-selective channel for example, the energy-optimal solution for WPT allocates power only to the strongest sub-channel, while the rate-optimal solution for WIT is the well-known water-filling power allocation over more than one strong sub-channels in general \cite{2010:Grover}.

Another important objective of this paper is to design an efficient feedback mechanism in WPT networks. As shown in \cite{2015:Bi1}, to maximize the network lifetime, it is important for the ENs to have the knowledge of both channel state information (CSI) and battery state information (BSI), i.e., the residual battery levels of WDs. Specifically, the knowledge of the strong sub-channels can boost the energy transfer efficiency, and the knowledge of those close-to-outage WDs can help avoid their energy outage by timely charging. In practice, transmitting CSI and BSI feedbacks may consume non-trivial amount of energy of the WDs and leave less time for WPT. Therefore, efficient CSI/BSI feedback is needed to maximize the net energy gain, i.e., the energy gain obtained from more refined charging control less by the feedback energy cost.

Our main contributions in this paper are as follows.
\begin{itemize}
  \item We propose a voting-based distributed charging control framework for broadband WPT networks. Specifically, each WD simply estimates the frequency sub-channels, casts its vote(s) for some strong sub-channel(s) and sends to the ENs along with its battery state, based on which each EN allocates its transmit power over the sub-channels independently. The proposed feedback method is low in complexity and applicable to practical WDs (e.g., RFID tags) only with simple baseband processing capability. Under the proposed framework, we study lifetime-maximizing CSI feedback and transmit power allocation designs.
  \item We derive the general expression of the expected lifetime achieved by a charging control method in WPT networks, which shows that a lifetime-maximizing charging control should be able to achieve a balance between the energy efficiency, user fairness and the induced energy cost of WPT. Some general principles are derived to guide the design of practical charging control method, e.g., the user priority-based charging scheduling.
  \item Based on the analysis, we propose practical power allocation algorithm with the considered voting-based CSI feedback. Specifically, the power allocated to a sub-channel is a function of the weighted sum vote received from all WDs, while the number of votes cast by a WD and the weight of each vote are related to its current energy level. Several effective power allocation functions are proposed. For practical implementation, we also discuss the setting of function parameters to maximize the network lifetime in practical systems.
  \end{itemize}
The network lifetime performance of the proposed distributed charging control methods is then evaluated through simulations under different setups. We show that the proposed voting-based charging control can effectively extend the network lifetime. Interestingly, we find that allocating all the transmit power of each EN to the best sub-channel that receives the highest vote achieves superior performance compared to other power allocation methods. In fact, this is consistent with the energy-optimal power allocation solution in point-to-point frequency-selective broadband channel, i.e., a special case of the multi-EN and multi-WD system considered in this paper. A related work in \cite{2014:Niyato} designs an interesting energy auction mechanism among the WDs in WPT networks to control the transmit power and shows the existence of an equilibrium. However, it only considers energy transfer on a narrowband channel instead of the frequency-selective broadband channel considered in this paper. Besides, the WDs are assumed selfish by nature and intend to harvest more energy. In our paper, however, we consider the WDs working collaboratively to achieve a common objective, e.g., monitoring the temperature of an area, such that a WD is not aimed to maximize its own harvested energy at the cost of reducing the lifetime of the whole network.

The rest of this paper is organized as follows. We first present in Section II a voting-based distributed charging control framework and and the key performance metric. In Section III, we analyze the expected network lifetime and derive the lifetime-maximizing design principles of wireless charging control. The detailed designs of power allocation and feedback mechanism are presented in Section IV. In Section V, simulation results are presented to evaluate the performance of the proposed charging control methods. Finally, the paper is concluded in Section VI.

\section{System Model}
\subsection{Channel Model}
\begin{figure}
\centering
  \begin{center}
    \includegraphics[width=0.45\textwidth]{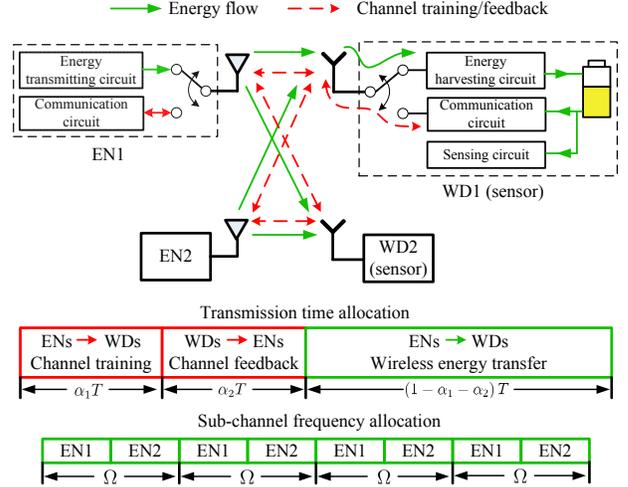}
  \end{center}
  \caption{An example system model of a broadband WPT-enabled sensor network, along with the transmission block time allocation and sub-channel allocation among the ENs.}
  \label{101}
\end{figure}

As shown in Fig.~\ref{101}, we consider a broadband WPT network, where $M$ ENs are connected to stable power sources and broadcast RF energy to power $K$ distributed WDs. The total bandwidth of the system and the channel coherence bandwidth are denoted by $D$ and $\Omega$, respectively, with $D \gg\Omega$. For simplicity, we assume that $D$ can be divided by $\Omega$ to form $N\triangleq D/\Omega$ parallel channels. To achieve full frequency diversity gain $N$ for each EN, each channel is further divided into $M$ sub-channels each for one of the $M$ ENs. The $N$ sub-channels allocated to the $i$-th EN are denoted by $\mathcal{E}_i$, $i=1,\cdots,M$, on which the EN can transmit narrowband energy signals. An example channel assignment is shown in Fig.~\ref{101}, where the adjacent sub-channels allocated to the same EN are separated by $\Omega$, thus the energy signals transmitted by the $i$-th EN to a WD experience independent fading over the $N$ sub-channels. Besides, the sub-channels of different ENs are also assumed to be independent due to sufficient spatial separations. We further assume that the wireless channels experience block fading, where the sub-channel gains remain constant in a transmission block of length $T$ and vary independently over different blocks.

For each WD, a single antenna is used for both energy harvesting and communication in a time-division-duplexing (TDD) manner (see WD1 in Fig.~\ref{101}). In particular, the communication circuit is used for channel estimation, i.e., receiving pilot signals sent by the ENs and sending channel feedback to the ENs. Besides, each WD may have a functional circuit to perform specific tasks, e.g., target sensing in Fig.~\ref{101}. For the $k$-th WD, the energy harvesting circuit converts the received RF signal to DC energy and store in a rechargeable battery of capacity $C_k$ to power the communication and functional circuits. On the other hand, each single-antenna EN also has a similar TDD circuit structure (see EN1 in Fig.~\ref{101}) to switch between energy transfer and communication with the WDs.

\subsection{CSI and BSI Feedback}
At the beginning of the $l$-th transmission block, $l=1,2,\cdots$, the $M$ ENs broadcast pilot signals simultaneously to the WDs in $\alpha_1 T$ time duration. Specifically, the $i$-th EN transmits pilot signals on the $N$ sub-channels in $\mathcal{E}_i$, $i=1,\cdots,M$. Upon receiving the pilot signals, each WD $k$ first estimates the $MN$ sub-channel gains, denoted by $h^l_{k,j}$, $j=1,\cdots,MN$. For the sub-channels in $\mathcal{E}_i$ allocated to the $i$-th EN, we assume that the channel gains from the EN to the $k$-th WD follow a general distribution with the equal mean given by
\begin{equation}
\label{1}
\mathbb{E}[h^l_{k,j}] = \beta d_{i,k}^{-\delta}, \ \ \forall j\in\mathcal{E}_i, \ l=1,2,\cdots,
\end{equation}
where $d_{i,k}$ denotes the distance between the $i$-th EN and the $k$-th WD, $\delta\geq 2$ denotes the path-loss exponent, and $\beta$ denotes a positive parameter related to the antenna gain and signal carrier frequency, which is assumed to be equal for all the sub-channels.

\begin{figure}
\centering
  \begin{center}
    \includegraphics[width=0.5\textwidth]{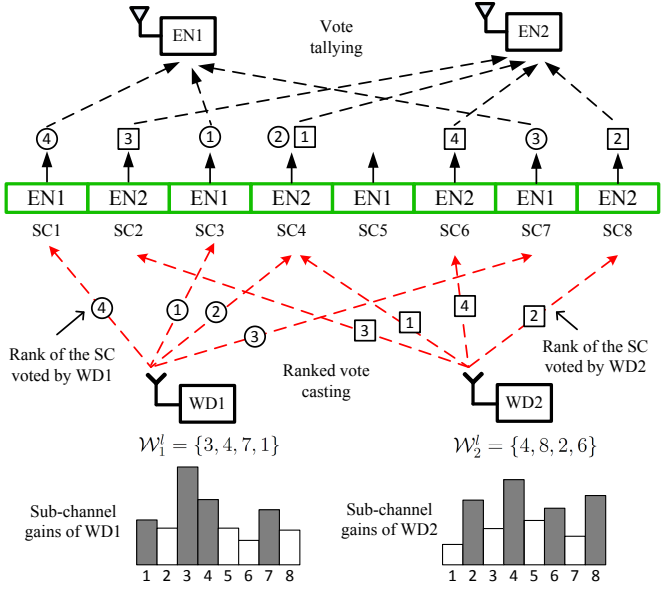}
  \end{center}
  \caption{Illustration of a voting-based channel feedback mechanism. Each of the WDs selects $4$ strongest sub-channels (SCs) and send their indices with ranks to the ENs.}
  \label{102}
\end{figure}

Then, the $K$ WDs feed back the channel gains to all the ENs in the next $\alpha_2 T$ time, which can be achieved either using orthogonal time slots or frequency bands. Conventional channel feedback procedure requires each WD to encode and modulate the $MN$ real channel gains, and send to the ENs. This, however, can be costly to the WDs due to some of the energy harvested consumed on channel feedback, or even infeasible due to the lack of adequate baseband processing capability of some simple energy-harvesting WDs. In light of this, we consider a practical voting-based feedback mechanism as shown in Fig.~\ref{102}. Specifically, each WD, say the $k$-th WD, simply estimates the received power levels of the $MN$ sub-channels, selects the $n_k^l$ strongest sub-channels, ranks them in a descending order based on the channel gains, and broadcasts the indices of the ordered $n_k^l$ sub-channels, denoted by $\mathcal{W}^l_k$, to the $M$ ENs. The rank of sub-channel $j\in \mathcal{W}^l_k$ is denoted by $R_{k,j} \in\left\{1,\cdots, n_k^l\right\}$. Notice that the value of $n_k^l$ is a design parameter to be specified later, which can be varying in different transmission block and across different WDs. For each EN, it observes the feedbacks from all the $K$ WDs, denoted by $\mathcal{W}^l \triangleq \left\{\mathcal{W}_1^l,\cdots,\mathcal{W}_K^l\right\}$. The channel feedback mechanism can be analogously considered as a voting system that the $k$-th elector (WD) casts $n_k^l$ \emph{ranked votes} for the $MN$ candidates (sub-channels).

Let $X_k^{l}$ denote the residual energy of the $k$-th WD at the end of $l$-th block, $E_k^{l}$ denote the amount of energy consumed within the block, including the energy spent on performing CSI feedback. For simplicity, we assume that the energy consumption rate is constant within each block, so that the energy level increases/decreases monotonically in each block. Then, the residual energy at the end of the $l$-th block is
\begin{equation}
\label{2}
X_k^{l} = \min\left\{\max\left(X_k^{l-1}-E_k^{l} + Q_k^{l},0\right),C_k\right\},\ \ l=1,2,\cdots,
\end{equation}
where $X_k^0$ denotes the initial energy level. In this paper, $E_k^{l}$ is assumed to follow a general distribution with an average consumption rate $\mathbb{E}[E_k^{l}]= \mu_k T$, $\forall l$. In particular, we assume $\mu_k \triangleq \bar{\mu}_k + \hat{\mu}_k$, where $\bar{\mu}_k > 0$ and $\hat{\mu}_k \geq 0$ denote the energy consumption unrelated and related to the energy harvesting performance, respectively. For instance, some WDs can perform transmit power control and CSI feedback rate variation adaptive to the instantaneous energy harvesting rate. In this paper, we only consider the impact of CSI feedback $\mathcal{W}^l$ to the WPT-related energy consumption rate $\hat{\mu}_k$, and do not consider other device energy management methods, e.g., transmit power control and device hibernation.

For simplicity, we assume that all the WDs have the same battery capacity, i.e., $C_k=C,\ \forall k$, and the battery capacity $\left[0,C\right]$ is divided into $I$ intervals specified by the thresholds $\left\{b_0, b_1,\cdots, b_{I-1}, b_{I}\right\}$, where $b_0=0$, $b_I=C$ and $b_i<b_j$ if $i<j$. We use $B_k^l$ to denote the battery state of WD $k$ at the end of the $l$-th transmission block, where the WD is referred to as in the $r$-th battery state, i.e., $B_k^l = r$, if the residual energy $X_k \in \left(b_{r-1},b_{r}\right]$, $r = 1, \cdots, I$. We assume that the WDs feed back $B_k^l$'s using a separate channel other than the one used for WPT. The ENs can keep a record on the battery states of the WDs, so that a WD $k$ only needs to broadcast a one-bit information indicating the change of battery state (to a lower or higher state) in the $l$-th transmission block. In this paper, we assume that all the WDs work collaboratively, such that each WD will report its true BSI to allow the ENs to make proper charging decisions to extend the network lifetime. In this case, the ENs have the knowledge of battery states of all the WDs at the beginning of the $l$-th transmission block, which is denoted by $\mathbf{B}^l = \left\{B_1^{l-1},\cdots,B_K^{l-1}\right\}$, $l=1,2,\cdots,$ and $B_k^0$ denotes the initial battery state of WD $k$. In practice, the one-bit BSI feedback is much infrequent than the CSI feedback, e.g., once several minutes versus several seconds, and has much less information to transmit, especially when $I$ is small. Therefore, we neglect the energy cost on BSI feedback in this paper.

\subsection{Transmit Power Allocation}
With both BSI ($\mathbf{B}^l$) and CSI ($\mathcal{W}^l$) feedbacks, the ENs allocate transmit power over the broadband channel in a distributed manner without the need of centralized control. It is worth mentioning that, although energy transfer can be performed on a narrow band, we exploit in this paper the frequency diversity gain in a multi-user environment to achieve more efficient and reliable energy transfer via transmit power allocation over multiple sub-channels. Besides, power allocation is only performed by the ENs to enhance the WPT performance. The communications between the WDs and the ENs are only for exchanging feedbacks and control signals for WPT, where no transmit power allocation for data transmission is considered.

The $i$-th EN, for instance, allocates its transmit power on the assigned sub-channels $\mathcal{E}_i$, denoted by $\left\{P^l_{j},\ \forall j\in \mathcal{E}_i \right\}$, where each EN has a total transmit power constraint $\sum_{j\in \mathcal{E}_i} P^l_{j} = P_0,\ i=1,\cdots,M,\ l=1,2,\cdots.$ In general, the power allocated by the $i$-th EN to the $j$-th sub-channel in the $l$-th transmission block can be expressed as a function of the available BSI and CSI:
\begin{equation}
\label{9}
P_j^l = f\left(\mathbf{B}^l,\mathcal{W}^l\right), \ j\in \mathcal{E}_i,\ l=1,2,\cdots.
\end{equation}
The design of power allocation function $f$ in (\ref{9}) will be discussed in detail in Section IV. Accordingly, the received energy by the $k$-th WD in the $l$-th transmission block is
\begin{equation}
\label{3}
Q_k^{l} = \eta \left(1-\alpha_1 -\alpha_2\right) T \cdot \mathsmaller\sum_{j=1}^{MN} P^l_{j} h^l_{k,j}, \ k=1,\cdots,K,
\end{equation}
where $\eta \in(0,1]$ is a fixed parameter denoting the energy harvesting efficiency and assumed equal for all WDs.

\subsection{Performance Metric}
The output voltage of a battery decreases with the residual energy level. We say an \emph{energy outage} occurs if the remaining energy level of a WD is below a certain threshold $\nu_k^{lo}$, such that normal device operation could not be maintained. Once a device is in energy outage, it is assumed to enter hibernation mode and become inactive. Given the initial battery level $\mathbf{X}^{0}=\left[X_1^0,\cdots,X_K^0\right]$, \emph{network lifetime} is defined as the duration until $\hat{K}$ out of the $K$ WDs are in energy outage, such that a network function achieved collectively by the $K$ WDs fails. For instance, the data reported by a sensor network is trustworthy when more than $K-\hat{K}$ sensors function properly, and considered unreliable otherwise. However, the network lifetime performance for the general $\hat{K}>1$ case is often analytically intractable due to the combinatorial nature of the WDs' operations. Like many previous studies on network lifetime (see e.g., \cite{2007:Chen}), we perform the analysis of charging control method for a special case of $\hat{K}=1$, i.e., a network reaches its lifetime as long as any WD is in outage. We will show by simulations later that a good design for the case of $\hat{K}=1$ also yields superior network lifetime performance for the general cases of $\hat{K}>1$. For the simplicity of illustration, we assume that $\nu_k^{lo} =0, \ \forall k$, throughout this paper. Then, the WDs are different only by their channel and energy consumption distributions.

Given the locations of the ENs and WDs, we could see from (\ref{2}) and (\ref{3}) that the harvested energy of the WDs, thus the network lifetime, is directly related to the transmit power allocation strategy $\mathbf{P}^{l}=[P_1^l,\cdots,P_{MN}^l]$ over $MN$ frequency sub-channels and time block $l=1,2,\cdots$. Meanwhile, we also notice from (\ref{1}) that the network lifetime is closely related to the locations of the ENs. In particular, the placement optimization of ENs has been studied in wireless powered communication networks where the locations of the WDs are fixed \cite{2015:Bi1}. In fact, the designs of transmit power allocation and EN placement are complementary to each other in different time-scales. That is, EN placement is designed in a large time-scale to deal with wireless signal path loss, while transmit power allocation is performed in a small time-scale to adapt to wireless channel fading and battery storage variation. In this paper, we assume that the placement of the ENs is given and focus on the design of lifetime-maximizing charging control method over channel and battery dynamics.

\section{Expected Lifetime of WPT Networks}
In this section, we analyze the impact of a charging control policy to the operating lifetime of WPT networks, defined as the duration until one of the WDs is in energy outage. In particular, we denote $L_\psi$ as the expected network lifetime achieved by a charging policy $\psi$, which specifies the transmit power allocation at each EN and in each transmission block, and thus determines the harvested energy $Q_k^{l}$, for $k=1,\cdots,K$ and $l=1,2,\cdots$. As a good charging policy should perform consistently regardless of the EN transmit power constraint $P_0$. To avoid trivial results, we assume that $P_0$ is sufficiently small, such that the expected network lifetime is finite regardless of the charging policy used, i.e.,
\begin{equation}
\underset{\psi \in \pi}{\text{maximize  }}L_\psi < \infty,
\end{equation}
where $\pi$ is the set of all feasible policies that satisfy the transmit power constraints. That is, the total energy harvesting rate of all the WDs is always lower than the total charging-independent consumption rate, i.e., $\sum_{k=1}^K\mathbb{E}[Q_k^{l}/T] < \sum_{k=1}^K \bar{\mu}_k$. In fact, simulation results in Section V find that a charging policy $\psi$ that achieves a longer $L_\psi$ under a small $P_0$ also requires lower transmit power to achieve nearly-perpetual network operation (i.e., $L_\psi$ is a very large number). Therefore, the study of a lifetime-maximizing charging policy under finite network lifetime assumption also has important implication in practical system designs with higher transmit power.

\subsection{Wireless Charging as Repeated Bets}\label{sec:networklifetime}
\begin{figure}
\label{model2}
\centering
  \begin{center}
    \includegraphics[width=0.5\textwidth]{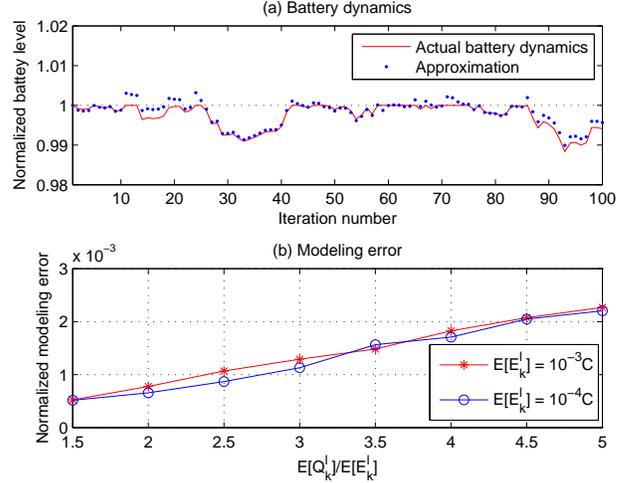}
  \end{center}
  \caption{Accuracy of the approximated battery dynamics. The figure (a) above: the actual v.s. approximated battery dynamics of the $k$-th WD when $\mathbb{E}[Q_k^l] = 1.5 \mathbb{E}[E_k^l]$ and $\mathbb{E}[E_k^l] = 10^{-3}C$; (b) below: the modeling error of net energy harvesting rate normalized against $\mathbb{E}[E_k^l]$ under different $\mathbb{E}[Q_k^l]/\mathbb{E}[E_k^l]$ ratios. Here, both $Q_k^l$'s and $E_k^l$'s are i.i.d. exponential variables for $l=1,2,\cdots$. }
  \label{111}
\end{figure}

The exact battery dynamic in (\ref{2}) complicates the analysis of network lifetime because of the max/min operators, yet failing to provide extra insight into charging policy design. To capture the essence of the battery dynamics, we make the following modifications:
\begin{itemize}
  \item the residual energy of a WD at the end of a transmission block could be negative when an energy outage occurs;
  \item the energy level could be higher than the battery capacity at the end of a transmission block, but such an overcharged battery cannot harvest any energy in the following transmission blocks until the energy level drops below the capacity at the end of a block.
\end{itemize}
The first modification overestimates the energy consumption in the last transmission block before the network reaches its lifetime, which has marginal effect on the modeling accuracy as the energy consumed within a transmission block is much smaller than the battery capacity. For the second modification, it has little impact to those WDs whose energy harvesting rates are smaller than or equal to the consumption rates ($\mathbb{E}[Q_k^l]\leq \mathbb{E}[E_k^l]$), e.g., WDs far away from the ENs, as they are rarely over-charged. Instead, it will impact the battery dynamics of close-to-EN WDs with $\mathbb{E}[Q_k^l]>\mathbb{E}[E_k^l]$. Specifically, it overestimates the energy harvested in a transmission block when a WD reaches its capacity, while it also underestimates the energy harvested in the future blocks before the energy level drops below the capacity. To better visualize the overall impact, we show in Fig.~\ref{111}(a) the actual and approximated battery levels over time for a WD with $\mathbb{E}[Q_k^l] = 1.5 \mathbb{E}[E_k^l]$ and $\mathbb{E}[E_k^l] = 10^{-3}C$. Besides, we also plot in Fig.~\ref{111}(b) the average modeling error in the sense of net energy harvested. We can see that the approximated battery dynamic in general overestimates the actual battery level. In particular, the modeling error is less than $0.3\%$ under different setups, indicating that the modified model can well approximate the actual energy harvesting process in the long term.

With the two modifications above, we could eliminate the max and min operators in (\ref{2}) and express the battery dynamics by a simple random process as follows
\begin{equation}
X_k^{l+1} = X_{k}^{l} - E_{k}^{l} + \hat{Q}_{k}^{l}, \ \ l=0,1,\cdots,
\end{equation}
where
\begin{equation}
\hat{Q}_{k}^{l}= \begin{cases}
0, & X_{k}^{l}\geq C,\\
Q_{k}^l, & \text{otherwise},
\end{cases}
\end{equation}
and $Q_{k}^l$ is given in (\ref{1}). In this case, the network fails as long as $X_{k}^{l}\leq 0$ for any $k$. Notice that both $Q_{k}^l$ and $E_{k}^{l}$ are related to the control policy $\psi$ in use, e.g., adaptive CSI feedback affects $E_{k}^{l}$. For the simplicity of exposition, we do not use different notations to indicate that $Q_{k}^l$ and $E_{k}^{l}$ are achieved by a specific policy $\psi$ in the following discussions.

Equivalently, the wireless charging process could be modeled as a group betting process with the $K$ WDs as gamblers. In particular, $X_k^{l}$ is the balance of gambler $k$, who repeatedly bets with a casino with $E_{k}^{l}$ as the income and $\hat{Q}_{k}^{l}$ as the loss in the $l$-th bet. The bet starts with each gambler $k$ holding $X_k^0$ initial balance, and stops once a gambler's balance becomes zero or negative. Then, the stopping time of the bet is also the network lifetime of the WPT network. Evidently, it is not a \emph{fair bet} because the average income and loss of each bet are not equal in general, i.e., $\mathbb{E}[E_{k}^{l}] \neq \mathbb{E}[\hat{Q}_{k}^{l}]$ for each $l$. In the following, we construct a fair game and derive the expected network lifetime using the \emph{Martingale} stopping time theorem \cite{2001:Grimmett}.

\subsection{Expected Network Lifetime}
The key idea of constructing a fair bet is to compensate the gamblers in each bet. We define a random process $\mathbf{Z}_l = [Z_1^{l},Z_2^{l},\cdots, Z_K^{l}]$, $l=0,1,\cdots$, with $Z_{k}^l = X_{k}^l + Y_{k}^l$, and
\begin{equation}
\label{15}
\small
Y_k^l= \begin{cases}
0, & l=0,\\
Y_k^{l-1} +  \mathbb{E}\left[E_{k}^{l} \mid \mathbf{B}^l\right] - \mathbb{E}\left[Q_{k}^{l}\mid \mathbf{B}^l\right], & l>0, X_{k}^{l-1} < C,  \\
Y_k^{l-1} + \mathbb{E}\left[E_{k}^{l} \mid \mathbf{B}^l\right],  &   l>0, X_{k}^{l-1} \geq C.
\end{cases}
\end{equation}
Here, $\mathbb{E}\left[Q_{k}^{l}\mid \mathbf{B}^l\right]$ denotes the average amount of energy received by the $k$-th WD in the $l$-th transmission block given that the WDs are in energy states $\mathbf{B}^l$ at the beginning of the transmission block, where the average is taken over the realizations of wireless channel fading of all the sub-channels in the $l$-th transmission block. Similarly, $\mathbb{E}\left[E_{k}^{l} \mid \mathbf{B}^l\right]$ denotes the average amount of energy consumed by the $k$-th WD conditioned on the current battery states. In particular, $Y_k^l$ could be considered as the cumulative compensation given to the gambler $k$ at the end of the $l$-th bet, where it is compensated for $\left( \mathbb{E}\left[E_{k}^{l} \mid \mathbf{B}^l\right] - \mathbb{E}\left[Q_{k}^{l}\mid \mathbf{B}^l\right]\right)$ in a bet if its balance is below $C$ in the previous bet and $\mathbb{E}\left[E_{k}^{l} \mid \mathbf{B}^l\right]$ otherwise. The following result shows that the random process $\mathbf{Z}_l$ is a Martingale.

\emph{Lemma 1:} The random process $\left\{\mathbf{Z}_l,l\geq 0\right\}$ is a Martingale, or equivalently the bet is a fair.

\emph{Proof:} To prove Lemma $1$, we need to show that for all $l$ it satisfies 1) $\mathbb{E}\left[Z^l_k\right] <\infty$, $\forall k$ and 2) $\mathbb{E}[\mathbf{Z}_{l+1} | \mathbf{Z}_l = \mathbf{z}_l, \cdots, \mathbf{Z}_1 =$ $\mathbf{z}_1] = \mathbf{z}_l$ \cite{2001:Grimmett}. Condition $1$) holds from the implicit assumption that the number of bets is finite. For condition $2$), we have for each $k$
\begin{equation*}
\begin{aligned}
&\mathbb{E}\left[Z_{k}^{l+1}\big|Z_k^l = z_k^l, \cdots, Z_k^0 = z_k^0\right] \\
=\ &z_k^l + \mathbb{E}\left[\hat{Q}_{k}^{l} - E_{k}^{l} \mid \mathbf{B}^l\right] + \mathbf{1}_{kl}^{C}\cdot \mathbb{E}\left[E_{k}^{l} \mid \mathbf{B}^l\right] \\
&+ \left(1- \mathbf{1}_{kl}^{C}\right)\cdot\left( \mathbb{E}\left[E_{k}^{l} \mid \mathbf{B}^l\right] - \mathbb{E}\left[Q_{k}^{l}\mid \mathbf{B}^l\right]\right)\\
= &\ z_k^l + \mathbf{1}_{kl}^{C} \left(\mathbb{E}\left[E_{k}^{l} \mid \mathbf{B}^l\right] - \mathbb{E}\left[E_{k}^{l} \mid \mathbf{B}^l\right] \right) + \left(1- \mathbf{1}_{kl}^{C}\right)\\
&\ \left(\mathbb{E}\left[Q_k^l-E_k^l\mid \mathbf{B}^l\right] +  \mathbb{E}\left[E_{k}^{l} \mid \mathbf{B}^l\right] - \mathbb{E}\left[Q_{k}^{l}\mid \mathbf{B}^l\right] \right) \\
=& \ z_k^l,
\end{aligned}
\end{equation*}
where $\mathbf{1}_{kl}^{C}$ is an indicator function that equals $1$ if $X_k^l\geq C$ and $0$ otherwise. This completes the proof.  $\hfill \blacksquare$

Then, the following Martingale Stopping Theorem \cite{2001:Grimmett} could be used to derive the expected network lifetime.

\emph{Proposition 1 (Martingale Stopping Theorem):} Let $\left\{\mathbf{Z}_l,l\geq 0\right\}$ be a Martingale and $W$ a stopping time that depends only on the value of $\mathbf{Z}_l$. If $\mathbb{E}\left[|z^W_k|\right]<\infty$, $\forall k$, then $\mathbb{E}\left[\mathbf{Z}_W\right]=\mathbb{E}\left[\mathbf{Z}_0\right]$.

In our problem, $W$ corresponds to the number of bets until $X_k^W\leq 0$ for some $Z_k^W$. Based on Proposition $1$, we have
\begin{equation}
\label{14}
\begin{aligned}
&\mathbb{E}\left[\sum_{k=1}^K Z_k^0\right] = \sum_{k=1}^K x_k^0 = \mathbb{E}\left[\sum_{k=1}^K Z_k^W\right]\\
=& \ \mathbb{E}\left[\sum_{k=1}^K X_k^W\right] + \mathbb{E}\left[\sum_{k=1}^K Y_k^W\right].
\end{aligned}
\end{equation}
Let $\varepsilon_0 \triangleq \sum_{k=1}^K x_k^0$ and $\varepsilon_r \triangleq \mathbb{E}\left[\sum_{k=1}^K X_k^W\right]$ denote the initial total energy and the expected total residual energy when outage occurs, we have
\begin{equation}
\label{19}
\varepsilon_0  - \varepsilon_r  =  \mathbb{E}\left[\sum_{k=1}^K Y_k^W\right].
\end{equation}

We consider $N$ independent experiments of the repeated betting process, where $N$ is sufficiently large. By the law of large numbers, it holds that
\begin{equation}
\label{16}
\lim_{N\rightarrow \infty}\frac{1}{N} \sum_{i=1}^N \sum_{k=1}^K  Y_{k}^{W_i} \rightarrow \mathbb{E}\left[\sum_{k=1}^K Y_k^W\right],
\end{equation}
where $W_i$ is the stopping time of the $i$-th experiment. By substituting (\ref{15}) into (\ref{16}), the LHS of (\ref{16}) can be further expressed as
\begin{equation}
\label{17}
\begin{aligned}
&\lim_{N\rightarrow \infty}\frac{1}{N} \sum_{i=1}^N \sum_{k=1}^K  Y_{k}^{W_i} \\
=&  \sum_{k=1}^K \biggl\{\lim_{N\rightarrow \infty}\frac{1}{N}\sum_{i=1}^N   \sum_{l=1}^{W_i} \mathbb{E}\left[E_{k}^{i,l} \mid \mathbf{B}^{i,l}\right]\\
& - \lim_{N\rightarrow \infty} \frac{1}{N}\sum_{i=1}^N \sum_{l=1}^{W_i}  \left(1 - \mathbf{1}_{ilk}^{C}\right)\mathbb{E}\left[Q_{k}^{i,l} \mid \mathbf{B}^{i,l}\right]\biggr\},
\end{aligned}
\end{equation}
where the superscript $i$ of $\left\{E_{k}^{i,l}, Q_{k}^{i,l},\mathbf{B}^{i,l}\right\}$ denotes the corresponding value in the $i$-th experiment. $\mathbf{1}_{ilk}^{C}$ denotes an indicator function that equals $1$ if $X_k^l\geq C$ in the $i$-th experiment and $0$ otherwise. In particular, the first term in the RHS of (\ref{17}) can be equivalently written as
\begin{equation}
\label{20}
\begin{aligned}
&\lim_{N\rightarrow \infty}\frac{1}{N}\sum_{i=1}^N   \sum_{l=1}^{W_i} \mathbb{E}\left[E_{k}^{i,l} \mid \mathbf{B}^{i,l}\right]\\
= &\lim_{N\rightarrow \infty}\frac{\sum_{i=1}^N W_i}{N}  \cdot \frac{\sum_{i=1}^N   \sum_{l=1}^{W_i} \mathbb{E}\left[E_{k}^{i,l} \mid \mathbf{B}^{i,l}\right]}{\sum_{i=1}^N W_i} \\
\triangleq & \ \mathbb{E}\left[W\right] \mathbb{E}\left[E_{k}\right],
\end{aligned}
\end{equation}
where $\mathbb{E}\left[W\right]$ denotes the average stopping time, and $\mathbb{E}\left[E_{k}\right]$ denotes the mean energy consumption of WD $k$ in a transmission block averaged over all the realizations of battery state $\mathbf{B}$. Similarly, the second term in the RHS of (\ref{17}) can be written as
\begin{equation}
\label{18}
\begin{aligned}
&\lim_{N\rightarrow \infty} \frac{1}{N}\sum_{i=1}^N \sum_{l=1}^{W_i}  \left(1 - \mathbf{1}_{ilk}^{C}\right)\mathbb{E}\left[Q_{k}^{i,l} \mid \mathbf{B}^{i,l}\right]\\
= & \lim_{N\rightarrow \infty}\frac{\sum_{i=1}^N W_i}{N} \cdot \Biggl(\frac{\sum_{i=1}^N \sum_{l=1}^{W_i} \mathbb{E}\left[Q_{k}^{i,l} \mid \mathbf{B}^{i,l}\right]  }{\sum_{i=1}^N W_i} \\
&\ \ - \frac{\sum_{i=1}^N \sum_{l=1}^{W_i}\mathbf{1}_{ilk}^{C}\mathbb{E}\left[Q_{k}^{i,l} \mid \mathbf{B}^{i,l}\right]}{\sum_{i=1}^N W_i}\Biggr)\\
\triangleq &\ \mathbb{E}\left[W\right]\cdot\left(\mathbb{E}\left[Q_k\right] - \mathbb{E}\left[Q_k^C\right]\right).
\end{aligned}
\end{equation}
where $\mathbb{E}\left[Q_k\right]$ denotes the mean energy transferred to the $k$-th WD in a transmission block averaged over all the battery states $\mathbf{B}$, and $\mathbb{E}\left[Q_k^C\right]$ denotes the average amount of energy transferred to the $k$-th WD, which, however, cannot be harvested by the WD because of battery over-charge, i.e., battery level is larger than or equal to the capacity. For the simplicity of exposition, we denote $\alpha_k \triangleq \mathbb{E}\left[Q_k^C\right]/\mathbb{E}\left[Q_k\right]$ as the portion of energy unable to be harvested by the $k$-th WD.

By substituting (\ref{20}) and (\ref{18}) into (\ref{19}), we have
\begin{equation}
\varepsilon_0  - \varepsilon_r = \mathbb{E}\left[W\right] \sum_{k=1}^K \left\{ \mathbb{E}\left[E_{k}\right] - \left(1- \alpha_k\right)\mathbb{E}\left[Q_k\right] \right\}
\end{equation}
Then, the expected waiting time conditioned on $\varepsilon_0$ is
\begin{equation}
\label{4}
\begin{aligned}
&\mathbb{E}\left[L|\varepsilon_0\right]=\mathbb{E}\left[WT\right] \\
 =& \ \frac{\varepsilon_0 - \varepsilon_r }{\sum_{k=1}^K  \mathbb{E}\left[E_{k}\right]/T - \sum_{k=1}^K\left(1- \alpha_k\right)\mathbb{E}\left[Q_k\right]/T } \\
 \triangleq &\ \frac{\varepsilon_0 - \varepsilon_r }{\sum_{k=1}^K  \mu_k - \sum_{k=1}^K\left(1- \alpha_k\right)\lambda_k},
\end{aligned}
\end{equation}
where $\lambda_k$ and $\mu_k$ denote respectively the average power transferred to and consumed by the $k$-th WD. We notice that $E\left[L|\varepsilon_0\right]$ is always positive by assumption, as the total energy harvesting rate is smaller than the consumption rate, i.e., $\sum_{k=1}^K \lambda_k <\sum_{k=1}^K \bar{\mu}_k \leq \sum_{k=1}^K \mu_k$.

\subsection{Charging Policy Analysis}\label{sec:analysis}
It is worth mentioning that the network lifetime expression in (\ref{4}) assumes no specific setups, e.g., the number of ENs or wireless channel distribution, thus is applicable to any general WPT network. To prolong the network lifetime in (\ref{4}), a charging policy should produce
\begin{enumerate}
  \item high effective energy harvested by the WDs, i.e., $\sum_{k=1}^K\left(1-\alpha_k\right)\lambda_k$;
  \item low total residual energy upon energy outage $\varepsilon_r$;
  \item low total energy consumption rates $\sum_{k=1}^K \mu_k$.
\end{enumerate}
For condition $1$), the ENs should maximize the \emph{energy efficiency} of wireless energy transfer, i.e., the energy received by the WDs less by that wasted due to overcharging. Therefore, a good charging policy should transfer as much energy as possible to the WDs given that their current batteries are not fully charged. This indicates that the ENs should assign lower priority to transmit energy to the WDs that are close-to-capacity.

However, maximizing energy efficiency does not translate to the low total residual energy $\varepsilon_r$ upon outage as required in condition $2$). Intuitively, suppose that a tagged WD is close-to-outage, maximizing the total energy received by the WDs may overlook the emergent energy requirement of the tagged WD, such that the large amount of energy harvested by the WDs of moderate/high energy levels will translate to higher $\varepsilon_r$ if the tagged WD dies out in the following transmission blocks due to the low energy harvesting rate. Recall that $\sum_{k=1}^K \lambda_k \leq \sum_{k=1}^K \mu_k$ holds, the average total residual energy of the WDs decreases as the time elapses. Therefore, to reduce $\varepsilon_r$, the ENs should give priority to charging those close-to-outage WDs to avoid imminent energy outage, which in fact advocates \emph{energy fairness} among the WDs. The ideal case is for all the WDs to drain their batteries simultaneously right before outage, i.e., $\varepsilon_r=0$.

For the last condition, the charging control design only affects $\mu_k$'s through designing the CSI feedback mechanism $\left\{\mathcal{W}_k^l, k=1,\cdots,K\right\}$ over time $l=1,2,\cdots$. Evidently, there is a design tradeoff in the amount of CSI feedback. In general, setting larger $n_k^l$'s, i.e., feeding back on more sub-channels, could allow the ENs to have a better estimation of the sub-channel conditions, and thus better power allocation decisions. However, this also induces higher \emph{energy cost} on transmitting the feedback signals, which can eventually offset the energy gain. Therefore, we need to carefully design CSI feedback to maximize the net energy gains of the WDs.

To sum up, a lifetime-maximizing charging control policy should be able to balance between energy efficiency, fairness and the induced energy cost. Specifically, it should follow the design principles listed below to control the power transfer in a transmission block:
\begin{itemize}
  \item[a)] assign higher priority to charging WDs that are close-to-outage, if any; and assign lower priority to charging WDs that are close-to-capacity, if any;
  \item[b)] maximize the total amount of energy transferred to the WDs under the assigned priorities;
  \item[c)] set proper amount of CSI feedback to maximize the net energy gains for the WDs.
\end{itemize}
In practical WPT networks, the above mentioned terms, such as ``close-to-outage" and ``priority", should be translated to realistic design parameters depending on the specific system setups, such as channel coherence bandwidth, transmit power limit and user energy consumption rate. In the next section, we apply the above design principles to study the transmit power allocation problem under the voting-based charging control framework introduced in Section II.

\section{Voting-based Distributed Wireless Charging Control}\label{sec:protocol}
In this section, we propose a voting-based distributed charging control policy, which includes the methods to assign weights to the votes, tally votes and allocate transmit power over frequency. We also propose a low-complexity protocol and discuss the practical design issues.

\subsection{Weight Assignment of Votes}
For convenience of exposition, we drop the superscript $l$ in all notations as the index of the transmission block, and focus on one particular transmission block. Recall that each EN $i$ is aware of the BSI $\mathbf{B}$ and (partial) CSI $\mathcal{W}$ from the voting-based feedback. Each EN $i$ can tally the votes to the sub-channels in $\mathcal{E}_i$, from which it can have a rough estimation of the EN-to-WD channel conditions and allocate the transmit power. Intuitively, a sub-channel should be allocated with more transmit power if it gets many high-ranked votes, because this indicates that larger total energy can be transferred to the WDs that share the same strong sub-channel (see principle $b$ in Section III.C). This implies that each vote should be weighted by the rank of vote among all the votes cast by the WD. Besides, to reflect on the design principle $a$) in Section III.C, the weight of a vote should be higher (or lower) if the WD casts the vote is close-to-outage (or close-to-capacity). As for the principle $c)$ in Section III.C, the number of votes cast by each WD should be reduced (or increased) whenever energy conservation is necessary (or not urgent).

From the above discussion, the weight assignment of the votes can be achieved through designing a weighting matrix $\mathbf{W} \in \mathbb{R}^{I \times J}_{+}$, where $I$ is the number of battery states and $J$ denotes the maximum number of votes any WD can cast, i.e., a WD can feed back at most $J$ channel indices. In particular, the number of votes that a WD $k$ can cast in any transmission block is determined by its current energy state $B_k$. Each entry $W_{i,j}$ indicates the positive weight of a vote if a WD that casts the vote is in the $i$-th battery state and the vote is ranked the $j$-th among all the votes cast by this WD. An example matrix $\mathbf{W}$ is shown as below,
\begin{equation}
\label{5}
\mathbf{W} = \left(
  \begin{array}{ccc}
    63 & 27 & 0 \\
    21 & 9 & 0  \\
    6 & 3& 1 \\
    1 & 0 & 0 \\
  \end{array}
\right).
\end{equation}
Here, we consider $I =4$ battery states, and assume that a WD in energy states $\{1,2,3,4\}$ feeds back $\{2,2,3,1\}$ sub-channel indices, respectively. Notice that some entries can be set as zero, e.g., $W_{1,3}$ and $W_{4,2}$. With $\mathbf{W}$ given in (\ref{5}), the vote cast by WD $k$ with rank $r$ has a weight $W_{B_k,r}$. For instance, a vote ranked the $2$nd among the votes cast by the WD $k$ in battery state $B_k=1$ is assigned a weight $W_{1,2} =27$.

We can see that the weight assignment method discussed above is consistent with the general design principles of WPT control: the charging priority of a WD is achieved through assigning higher (lower) weight to its vote if the WD is in lower (higher) battery state; the charging efficiency is maximized through assigning higher (lower) weight to the entry in each row that corresponds to higher (lower) sub-channel gains; while WDs in different battery states can balance the energy gain and cost through feeding back different number of channel indices. The value of the weighting matrix $\mathbf{W}$ has direct effect to the performance of the WPT network, which will be discussed in Section IV.C. For the moment, we assume that $\mathbf{W}$ is known by all the ENs and study the associated transmit power allocation method in the next subsection.

\subsection{Transmit Power Allocation}
Following the general design principles, there are multiple ways to design the power allocation function $f\left(\mathbf{B},\mathcal{W}\right)$ in (\ref{9}), depending on the methods used to tally the votes and accordingly allocate the power over frequency. Here, we introduce two vote tallying methods and two power allocation methods, which can be combined to generate $4$ power allocation functions. Specifically, the two vote tallying methods are given as follows first.
\begin{enumerate}
  \item \emph{Universal Tallying:} Each EN tallies all the votes cast by the WDs. Specifically, based on the weighting matrix $\mathbf{W}$, each EN $i$ can compute the weighted sum vote to the $j$-th sub-channel in $\mathcal{E}_i$ as
\begin{equation}
\label{7}
v_j = \sum_{k=1}^K \mathbf{1}\left[j\in \mathcal{W}_k \right] \cdot W_{B_k,R_{j,k}}, \ j\in\mathcal{E}_i,
\end{equation}
where $R_{j,k}$ is the rank of sub-channel $j$ among the votes cast by WD $k$, and $\mathbf{1}\left[j\in \mathcal{W}_k \right]$ is an indicator function with value $1$ if $j\in \mathcal{W}_k$ and $0$ otherwise.
  \item \emph{Prioritized Tallying:} In this case, among the WDs that vote, each EN only tallies the votes cast by the WDs in the lowest battery state, i.e., the WDs currently with the highest priority. Suppose that among the WDs that cast votes to the sub-channels in $\mathcal{E}_i$, the WD(s) of the lowest battery state is (are) in the $p$-th battery state, where $p$ not necessarily equals $1$. Then, the weighted sum vote to the $j$-th sub-channel is given as
      \begin{equation}
      \label{10}
      v_j = \sum_{k=1}^K \mathbf{1}\left[B_k= p\right]\cdot \mathbf{1}\left[j\in \mathcal{W}_k \right] \cdot W_{B_k,R_{j,k}}, \ j\in\mathcal{E}_i,
      \end{equation}
      where $\mathbf{1}\left[B_k= p\right]$ is an indicator function with value $1$ if $B_k= p$ and $0$ otherwise.
\end{enumerate}

Furthermore, with either of the above two vote tallying methods, the following two power allocation strategies can be applied.
\begin{enumerate}
  \item \emph{Single-channel Allocation:} Allocate all the power to the sub-channel that receives the highest weighted sum vote. Specifically, the power allocated by the $i$-th EN to the $j$-th sub-channel is
      \begin{equation}
      \label{11}
      P_{j} = \begin{cases}
      P_0/N, & v_l = 0, \ \forall l\in\mathcal{E}_i,\\
      P_0, &  \exists v_l >0, l\in\mathcal{E}_i \text{  and  }j = \arg \max_{l\in\mathcal{E}_i} v_l,\\
      0, & \text{otherwise},
      \end{cases}
      \end{equation}
      where $v_l$ is given in (\ref{7}) or (\ref{10}). The first case corresponds to the scenario that the sub-channels in $\mathcal{E}_i$ receive no vote from the WDs. As the $i$-th EN has no knowledge of the current wireless channel conditions, it allocates equally the transmit power among the $N$ sub-channels in $\mathcal{E}_i$. Besides, if multiple sub-channels have the same weighted sum vote $v_l$, we randomly pick one and allocate all the transmit power to it.
  \item \emph{Proportional Allocation:} The transmit power of the $i$-th EN is allocated proportionally to the weighted sum vote received by each sub-channel in $\mathcal{E}_i$, i.e.,
      \begin{equation}
\label{12}
P_{j} = \begin{cases}
P_0/N, & v_l = 0, \ \forall l\in\mathcal{E}_i,\\
P_0 v_j/\sum_{l\in\mathcal{E}_i} v_l, & \text{otherwise}.
\end{cases}
\end{equation}
\end{enumerate}

The above vote tallying and power allocation methods can find their deep roots in real-life politics. On one hand, the universal tallying corresponds to the universal suffrage system that everyone's vote counts, while the prioritized tallying is analogous to parliament election system, where only the parliament members (prioritized voters) get to vote, rather than the common public. On the other hand, the single-channel power allocation is analogous to the winner-gets-all presidential election, while the proportional power allocation can be considered as the parliament election, where the number of seats that a party controls in the parliament is proportional to the votes it receives. In practice, each of the vote tallying methods can be flexibly combined with the power allocation methods. However, as it is an inconclusive question to real-life politics of which form of election method is the best, for the time being we do not have a conclusion about which combination is lifetime-maximizing in WPT networks. Instead, we address this question based on simulation results later in Section VI. Interestingly, we find by simulations that the single-channel power allocation achieves evident performance gain over the proportional power allocation, and the universal tallying can further improve the network lifetime performance.

\subsection{Protocol Description}
\begin{figure}
\centering
  \begin{center}
    \includegraphics[width=0.45\textwidth]{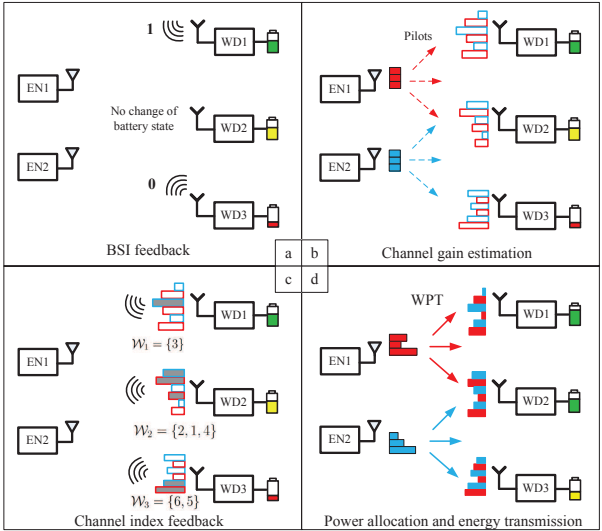}
  \end{center}
  \caption{Illustration of the voting-based distributed charging control protocol. Sub-figures (a)-(d) correspond to steps 1)-4) of the protocol
descriptions, respectively.}
  \label{103}
\end{figure}

In the following, we summarize the designs in this section as a voting-based distributed charging control protocol that operates in the following steps and is illustrated in Fig.~\ref{103}:
\begin{enumerate}
  \item At the beginning of each transmission block, each WD reports to the ENs a one-bit information indicating the change of battery state, if any, from which all the ENs know the BSI of all the WDs, i.e., $\left\{B_k, k=1,\cdots,K\right\}$;
  \item Each EN $i$ sends pilot signals on its $N$ sub-channels in $\mathcal{E}_i$, $i=1,\cdots, M$. Then, each WD $k$, estimates its own $MN$ sub-channel gains, denoted by $\hat{h}_{k,j}$'s for $j=1,\cdots,MN$.
  \item Each WD $k$ selects the $J_{B_k}$ strongest sub-channels from the $MN$ sub-channels by ordering $\hat{h}_{k,j}$, $j=1,\cdots,MN$, where $J_{B_k}$ is the number of non-zero entries in the $B_k$-th row of the weighting matrix $\mathbf{W}$. Then, each WD $k$ broadcasts the ordered indices of the $J_{B_k}$ sub-channels (i.e., $\mathcal{W}_k$).
  \item Based on $B_k$'s and $\mathcal{W}_k$'s, each EN independently allocates transmit power according to a combination of the vote tallying and power allocation methods introduced in Section IV.B. The WDs harvest RF energy in the remaining transmission block. Then, the iteration repeats from Step $1)$.
\end{enumerate}

The proposed charging control protocol incurs little signaling overhead exchanged between the ENs and the WDs. Specifically, each WD only needs to send out limited number of sub-channel indices based on the estimated channel gains and its own residual energy level, and broadcasts a simple one-bit BSI message only when its battery state changes. Besides, the protocol has low computational complexity and requires no coordination among the ENs. Each EN $i$ \emph{independently} tallies the received votes to the sub-channels in $\mathcal{E}_i$, and computes its own power allocation using simple power allocation function as in (\ref{11}) or (\ref{12}). The entries in $\mathbf{W}$ are the key design parameters of the proposed voting-based charging control protocol. A point to notice is that the value of $\mathbf{W}$ only needs to be determined once throughout the entire network operating lifetime. In particular, we can design the value of $\mathbf{W}$ in an offline manner and allow the ENs to inform $\mathbf{W}$ to all the WDs at the very beginning of the network operation. In this sense, the energy-limited WDs do not bear any computational complexity in the design of $\mathbf{W}$. In the next subsection, we have some discussions on the design of $\mathbf{W}$.

\subsection{Discussion on Weighting Matrix Design}
The design of $\mathbf{W}$ includes: 1) the number of rows $I$, i.e., the number of battery states (and the corresponding battery thresholds) of the WDs; 2) the number of non-zero entries in each row; and 3) the value of each non-zero entry $W_{i,j}$. In practice, setting the parameters of $\mathbf{W}$ is an art under specific network setup, however, still has some rules to follow as discussed below.

To begin with, using a larger number of battery states can improve the ENs' knowledge of the residual device energy levels, thus achieving more accurate charging priority assignment of the WDs. However, this also increases the frequency of the one-bit BSI feedback and accordingly the energy cost of the WDs. In practice, setting a small number of battery states, e.g., $I = 5$, would be sufficient to achieve satisfactory priority-based charging control. On the other hand, the thresholds of the battery states, i.e., $\left\{b_1,\cdots,b_{I-1}\right\}$, are not necessarily uniform. In fact, setting denser thresholds at low battery region can help ENs better identify the WD in the most urgent energy outage situation so as to arrange timely charging to it.

Secondly, the number of votes a WD casts is related to both the energy cost of sending a channel index feedback and its current battery state. The weighting matrix $\mathbf{W}$ in (\ref{5}) gives a good example to set the feedback amount of a WD in different battery states: the close-to-outage WDs should only send very few channel feedbacks to save energy. However, the number of feedbacks cannot be too small as well (e.g.,  $2$ votes in (\ref{5})), because more channel feedbacks allow multiple ENs to allocate more transmit power in favor of it; while those close-to-capacity WDs are currently not in need of energy transfer, and thus only need to cast one vote to indicate its strongest sub-channel; in between, the WDs of moderate battery levels should feed back several sub-channels (e.g., $3$ votes in (\ref{5})) to maximize the harvested energy without worrying too much about the cost on feedback or battery overcharging.

Finally, the values of non-zero $W_{i_1,j}$'s in the $i_1$-th row should be larger than $W_{i_2,j}$'s in the $i_2$-th row when $i_1<i_2$ to ensure that higher charging priority is given to WDs with lower residual battery. In (\ref{5}), for instance, the sum of the $1$st row is $2$ times larger than that in the $2$nd row, which is subsequently $2$ times larger than that in the $3$rd row. The rationale is that the number of close-to-outage WDs is often much smaller than those in moderate energy states. Setting a much higher value for the entries in the lower energy states can make sure the votes cast by the close-to-outage WDs are not overwhelmed by the many votes cast by the WDs in higher energy states. Within each row, a larger portion of the sum row weight should be given to the first entry, to increase the power allocated to the best sub-channel. Following the above discussions, the impact of $\mathbf{W}$ on the network lifetime is evaluated by simulations in the next section.

\begin{table}
\caption{Simulation Parameters}
\footnotesize
\begin{center}
\begin{tabular}{|c|c||c|c|}
 \hline
  EN Tx power &   $1$ W & Path loss exponent &   $2$\\ \hline
  Central frequency &   $915$ MHz &  Tx block length &   $500$ ms \\ \hline
  No. of SCs &   $30$  & Ave. WD power &   $3$ mW\\ \hline
  SC bandwidth &   $10$ KHz   & Battery voltage &   $1$ V\\ \hline
  Tx antenna gain &   $2$  & Battery capacity ($C$)&   $1000$ mAh\\ \hline
  Rx antenna gain &   $2$  & Feedback power per SC &   $0.1$ mW\\ \hline
\end{tabular}
\end{center}
\label{stat}
\end{table}

\section{Simulation Results}
In this section, we evaluate the performance of the proposed voting-based charging control protocol. In all simulations, we use the Powercast TX91501-1W transmitter as the ENs and P2110 Powerharvester as the energy receiver at each WD with $\eta= 0.51$ energy harvesting efficiency. Unless otherwise stated, the simulation parameters are listed in Table I, which correspond to a typical indoor sensor network. The weighting matrix $\mathbf{W}$ is as given in (\ref{5}), where the threshold vector for the battery is $\left\{0,0.3C,0.5C,0.9C,C\right\}$. Besides, we consider a stochastic energy consumption model that a WD consumes $12$ mW power with probability $0.25$ within a block, and no power with probability $0.75$. In this case, the average power consumption rate is $3$ mW for each WD. We set the initial battery level of all WDs as $0.75C$, such that the battery will be depleted in about $250$ hours without WPT.

\begin{figure}
\centering
  \begin{center}
    \includegraphics[width=0.45\textwidth]{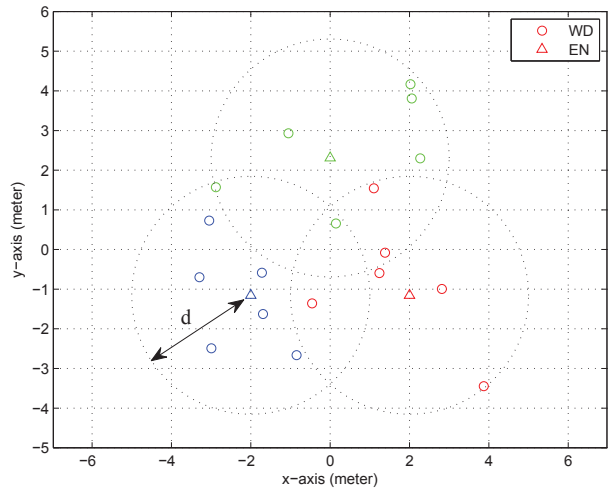}
  \end{center}
  \caption{An example placement of a WPT network with $3$ ENs and $18$ WDs.}
  \label{104}
\end{figure}

The wireless channel power gains follow exponential distributions with mean obtained from the path loss model. Without loss of generality, we consider a network fails if more than $K/3$ WDs are in energy outage. Unless otherwise stated, all simulations are performed in a simple $3$-EN WPT network shown in Fig.~\ref{104}, where the ENs are located at $\left\{(-2,-2/\sqrt{3}),(2,-2/\sqrt{3}),(0,4/\sqrt{3})\right\}$. In particular, $6$ WDs are randomly placed within a circle of radius $d$ centered at each EN, i.e., $K=18$ WDs. In general, a larger $d$ indicates a larger disparity among the users in the wireless channel conditions, and also a larger distance between the WDs to the ENs, which will translate to a shorter network lifetime in general.

For performance comparison, we consider the four power allocation functions from the combinations of the two vote tallying and two power allocation methods described in Section IV.B:
\begin{itemize}
  \item Singl-Univ: Single-channel power allocation based on universal vote tallying;
  \item Singl-Prio: Single-channel power allocation based on prioritized vote tallying;
  \item Propo-Univ: Proportional power allocation based on universal vote tallying;
  \item Propo-Prio: Proportional power allocation based on prioritized vote tallying.
\end{itemize}
Besides, we also consider five other representative benchmark schemes:
\begin{itemize}
  \item EqlPower: power is equally allocated to all the sub-channels by each EN;
  \item Singl-Unwt: the power of each EN is all allocated to the single sub-channel that receives the most number of votes, i.e., the votes are unweighted;
  \item Propo-Unwt: power is allocated proportionally to the number of unweighted votes that each sub-channel receives at each EN;
  \item Singl-Greedy: each greedy user votes for only the best sub-channel, and the power of each EN is all allocated to the single sub-channel that receives the most number of votes;
  \item Propo-Greedy: each greedy user votes for only the best sub-channel, and power is allocated proportionally to the number of votes that each sub-channel receives at each EN.
\end{itemize}
Because the EqlPower scheme is completely oblivious to CSI, we assume that $100\%$ of the time is used for WPT without any signaling overhead. For fair comparison, we assume that the other schemes use the same CSI feedback mechanism, where $\alpha_1 = 2\%$ of the time is spent on sending pilot signals, $\alpha_2 = 3\%$ of the time is spent on CSI feedback, and the rest $95\%$ is for WPT.

\begin{figure}
\centering
  \begin{center}
    \includegraphics[width=0.5\textwidth]{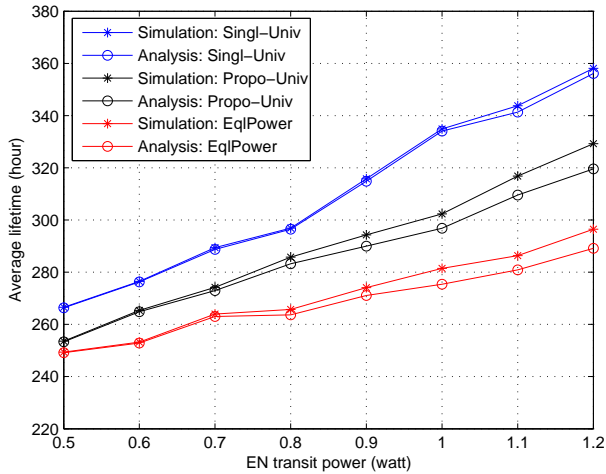}
  \end{center}
  \caption{Comparison of analysis and simulations of average network lifetime of three charging control schemes. }
  \label{110}
\end{figure}

\subsection{Analysis Validation}
We first verify the analysis of expected network lifetime expression derived in (\ref{4}). For the simplicity of exposition, we consider without loss of generality the EqlPower, Singl-Univ, and Propo-Univ schemes, and compare in Fig.~\ref{110} their average network lifetime by simulations and analysis under different EN transmit power. To be consistent with the analysis in Section III, we define that a network reaches its lifetime if any WD is in energy outage. We consider a specific realization of the random placement of the $18$ WDs in Fig.~\ref{104}. Each point in the figure is an average of $40$ independent simulations. We can see that all the analytical results are very close to the simulations. In general, the analysis underestimates the network lifetime, as the second modification of the battery dynamic overestimates the battery levels, thus leading to larger $\varepsilon_r$ in (\ref{4}). The analysis is especially accurate when the transmit power is small, and becomes less accurate as the transmit power increases because of the increase of over-charging probability. Overall, the average difference between the analysis and simulation is less than $1\%$ of the simulation value, which verifies the validity of our analysis in (\ref{4}).

\begin{figure}
\centering
  \begin{center}
    \includegraphics[width=0.5\textwidth]{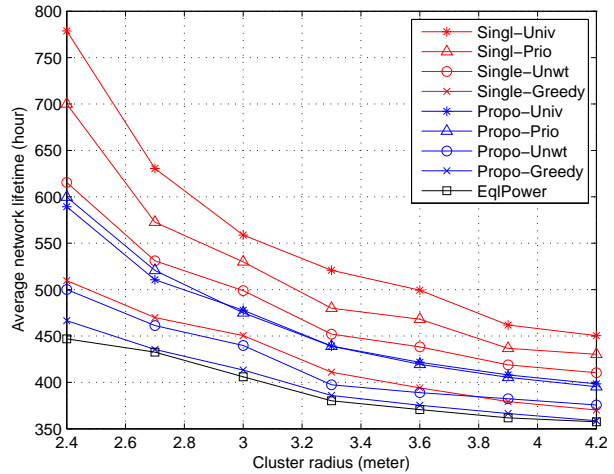}
  \end{center}
  \caption{Comparison of average network life achieved by different power allocation functions.}
  \label{105}
\end{figure}

\subsection{Lifetime Performance Comparison}
In Fig.~\ref{105}, we plot the average network lifetime achieved by different power allocation functions. Unless otherwise stated, each point in the figure is an average performance of $15$ random placements of the WDs, and the lifetime of a particular placement is an average of $10$ independent simulations over random wireless channels and device power consumptions. For all the schemes, the network lifetime decreases as $d$ increases, as expected. We can see that significant frequency diversity gain can be achieved from power allocation, where the channel-oblivious EqlPower scheme has the worst performance. Meanwhile, under the same vote-tallying method, a scheme that employs single-channel power allocation achieves evidently longer lifetime than using proportional power allocation. One explanation is that the single-channel power allocation can maximize the energy transferred to a particular WD in the current time slot, which is more effective to avoid energy outage for the WDs in urgent battery outage situations. Meanwhile, we can also see that each EN should tally the votes from all the WDs (instead of only the WDs in the lowest battery state), where Singl-Univ performs better than the Singl-Prio scheme. Besides, a WD should cast multiple votes when it is in need of energy, where the two greedy user schemes (Singl-Greedy and Propo-Greedy) perform poorly. In addition, the schemes using weighted votes (Singl-Univ and Singl-Prio) based on CSI and BSI feedbacks have much better performance than the one using unweighted votes (Singl-Unwt). In particular, the best-performing Singl-Univ method achieves on average $20\%$ longer lifetime than the Propo-Univ scheme, and over $40\%$ longer lifetime than the EqlPower scheme. The simulation results reveal an interesting finding in WPT networks that transmit power should be allocated to the best sub-channel. In fact, this is consistent with the energy-optimal power allocation solution in point-to-point frequency-selective channel, a special case of the multi-EN and multi-WD system considered. Besides, the selection of the best sub-channel should consider the votes from all the WDs.

\begin{figure}
\centering
  \begin{center}
    \includegraphics[width=0.5\textwidth]{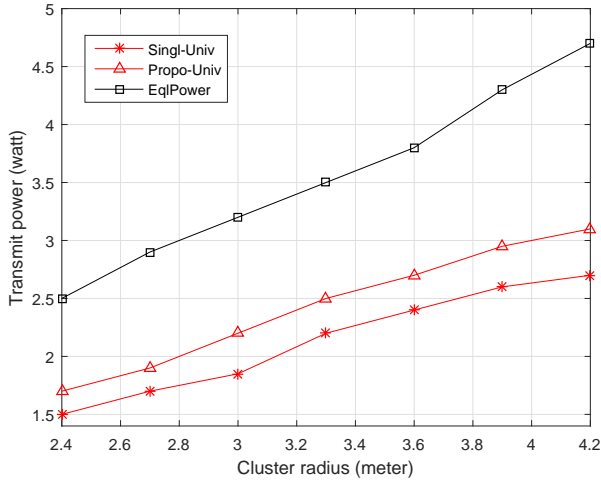}
  \end{center}
  \caption{Minimum transmit power of each EN required to achieve nearly-perpetual network operation.}
  \label{106}
\end{figure}

In Fig.~\ref{106}, we plot the minimum transmit power required by each EN to achieve nearly-perpetual network operation. For the simplicity of illustration, we consider three representative schemes: Singl-Univ, Propo-Univ, and EqlPower. Due to the randomness of channel fading and energy consumptions, it is not possible to truly sustain perpetual network operation. Here, a WPT system is said nearly-perpetual if the network lifetime is longer than $5000$ hours in all the $10$ independent simulations conducted. For each $d$, we randomly generate $5$ placements and calculate the average minimum transmit power required for each of the placements. The best-performing (worst-performing) Singl-Univ (EqlPower) scheme in Fig.~\ref{105} also require the lowest (highest) transmit power to achieve nearly-perpetual operation in Fig.~\ref{106}. In particular, the Singl-Univ scheme can save more than $40\%$ of the transmit power than that required by the EqlPower scheme. The results in Figs. \ref{105} and \ref{106} demonstrate the effectiveness of the proposed voting-based charging control method in extending the network lifetime, and shows that a scheme that achieves a longer network lifetime under low transmit power is in general also more power-efficient to achieve self-sustainable operation in practical WPT networks with higher power.

\subsection{Impact of Weighting Parameters}
\begin{figure}
\centering
  \begin{center}
    \includegraphics[width=0.5\textwidth]{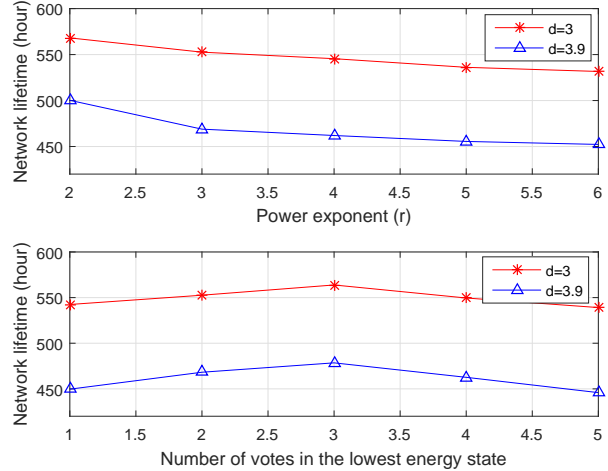}
  \end{center}
  \caption{Impact of weighting matrix in (\ref{13}) on network lifetime performance when: (a) the power exponent $r$ changes; or (b) the amount of CSI feedback amount changes.}
  \label{107}
\end{figure}

In this subsection, we use the best-performing Singl-Univ scheme to investigate the impact of the weighting matrix $\mathbf{W}$ on the system performance. In particular, we first examine the network lifetime when changing the value of $W_{i,j}$ for those $W_{i,j}\neq 0$. Specifically, we keep a fixed number of non-zero entries in $\mathbf{W}$ and only change the values of $W_{i,j}$'s. For the simplicity of illustration, we consider a weighting matrix as a function of power exponent $r>1$ as follows:
\begin{equation}
\label{13}
\mathbf{W}(r) = \left(
  \begin{array}{lll}
    7 r^2 & 3 r^2 & 0 \\
    7 r & 3 r & 0  \\
    6 & 3& 1 \\
    1 & 0 & 0 \\
  \end{array}
\right).
\end{equation}
Notice that the weight matrix $\mathbf{W}$ in (\ref{5}) corresponds to the case with $r=3$ in (\ref{13}). Evidently, a larger $r$ will lead to a larger difference of weights of the votes cast by WDs in different battery states. In Fig.~\ref{107}(a), we plot the average network lifetime as a function of $r$ when the cluster radius $d = 3$ or $3.9$ meters. We can see that the lifetime decreases when $r$ increase from $2$ to $6$. Intuitively, this is because assigning very large weights to the votes cast by WDs in lower battery states essentially approaches the worse-performing prioritized tallying method, where votes cast by WDs in higher battery states are neglected.  However, the simulation results in Fig.~\ref{107}(a) do not imply that a small weight is more favorable for the votes cast by WDs of low battery state. Instead, we can infer that the setting of $\mathbf{W}$ needs to balance between the energy efficiency and fairness among all the WDs.

At last, we investigate the performance tradeoff in terms of the CSI feedback amount, i.e., the number non-zero entries in $\mathbf{W}$. Specifically, we consider a $\mathbf{\hat{W}}$ of $4$ rows (i.e., $4$ fixed battery states) and varying number of columns (i.e., variable feedback amount). The non-zero elements in the $3$rd and $4$-th rows of $\mathbf{\hat{W}}$ are the same as those of $\mathbf{W}$ in (\ref{5}), while the non-zero elements in the first and the second rows are set as
\begin{equation}
\left\{90\right\},\{63,27\},\{63,27,9\},\{63,27,9,3\},\left\{63,27,9,3,1\right\}
\end{equation}
and
\begin{equation}
\{30\},\{21,9\},\{21,9,3\},\{21,9,3,1\},\{21,9,3,1,1\},
\end{equation}
respectively, in $5$ different feedback designs. That is, a WD casts $1$ vote when it is in the $4$-th battery state, $3$ votes in the $3$rd battery state, and a variable number of votes from $1$ to $5$ when it is in the $1$st or $2$nd battery states. The $\mathbf{W}$ in (\ref{5}) correspond to the case when the WDs cast $2$ votes in the first two battery states. The power consumed on transmitting each vote is $0.1$ mW. Besides, the time reserved on CSI feedback is assumed proportional to the maximum of non-zero entries among the $4$ rows in $\mathbf{\hat{W}}$. For instance, when the feedback number is $2$ for WDs in the $1$st or $2$nd battery state, the CSI feedback occupies $\alpha_2 = 3\%$ of a transmission block, because a WD casts $3$ votes at maximum when it is in the $3$rd battery state; when the maximum feedback number is $5$, however, we have $\alpha_2 = 5\%$. The network lifetime performance under different feedback settings is shown in Fig.~\ref{107}(b) when $d=3$ or $3.9$ meters. We can see that the network lifetime increases when the feedback amount increases from $1$ to $3$, indicating that the energy gain obtained from more refined CSI feedback outweighs the extra energy cost on sending more CSI feedbacks. However, as we further increase the feedback amount, the network lifetime decreases mainly because of the extra time consumed on sending CSI feedback to the ENs, which leaves less time for WPT transmission in a transmission block. We can therefore infer that a proper feedback amount should be selected. Meanwhile, because the best-performing Singl-Univ scheme allocates transmit power to only one sub-channel for each EN, the feedback amount should be set small to increase the chance of the strongest sub-channel being selected for transmission and also reduce the energy cost due to less WPT time resulted.

\section{Conclusions}
In this paper, we proposed a voting-based distributed charging control framework in multi-EN broadband WPT networks, to exploit frequency diversity gain to maximize the network operating lifetime. The proposed voting-based channel feedback mechanism is especially suitable for wireless powered devices with simple hardware structure and stringent battery constraint. Under the proposed framework, we studied the power allocation method and efficient CSI/BSI feedback design over multiple sub-channels. In particular, we derived the expected network lifetime of general WPT networks to draw the guideline of designing practical lifetime-maximizing charging control policies, and proposed accordingly efficient voting-based power allocation schemes with battery-state dependent CSI feedbacks. The effectiveness of the proposed methods in extending the network lifetime has been verified by extensive simulations. Interestingly, we found that superior lifetime performance is achievable by allocating all the transmit power of each EN to the best sub-channel that receives the highest weighted sum vote from all the WDs, instead of spreading the transmit power over multiple sub-channels. Practical system design issues were also discussed and examined through simulations.


\begin{thebibliography}{1}
\small

\bibitem{2002:Biyikoglu}
E.~Uysal-Biyikoglu, B.~Prabhakar, and A.~El Gamal, ``Energy-efficient packet transmission over a wireless link," \emph{IEEE/ACM Trans. Netw.}, vol.~10, no.~4, pp.~487-499, Aug.~2002.

\bibitem{2004:Younis}
O.~Younis and S.~Fahmy, ``HEED: a hybrid, energy-efficient, distributed clustering approach for ad hoc sensor networks," \emph{IEEE Trans. Mobile Comput.}, vol.~3, no.~4, pp.~366-379, Oct.~2004.

\bibitem{2007:Chen}
Y.~Chen and Q.~Zhao, ``An integrated approach to energy-aware medium access for wireless sensor networks," \emph{IEEE Trans. Signal Processing}, vol.~55, no.~7, pp.~3429-3444, Jul.~2007.

\bibitem{2015:Bi}
S.~Bi, C.~K.~Ho, and R.~Zhang, ``Wireless powered communication: opportunities and challenges," \emph{IEEE Commun. Mag.}, vol.~53, no.~4, pp.~117-125, Apr. 2015.

\bibitem{2015:Lu}
X.~Lu, P.~Wang, D.~Niyato, D.~I.~Kim, and Z.~Han, ``Wireless networks with RF energy harvesting: a contemporary survey," \emph{IEEE Commun. Surveys Tuts.}, vol.~17, no.~2, pp.~757-789, 2015.

\bibitem{2016:Bi}
S.~Bi, Y.~Zeng, and R.~Zhang, ``Wireless powered communication networks: an overview," \emph{IEEE Wireless Commun.}, vol.~23, no.~2, pp.~10-18, Apr.~2016.

\bibitem{2015:Ulukus}
S.~Ulukus, A.~Yener, E.~Erkip, O.~Simeone, M.~Zorzi, P.~Grover, and K.~Huang, ``Energy harvesting wireless communications: a review of recent advances," \emph{IEEE J. Sel. Areas Commun.}, vol.~33, no.~3, pp.~360-381, Mar.~2015.

\bibitem{2014:Krikidis}
I.~Krikidis, S.~Timotheou, S.~Nikolaou, G.~Zheng, D.~W.~K.~Ng, and R.~Schober, ``Simultaneous wireless information and power transfer in modern communication systems," \emph{IEEE Commun. Mag.}, vol.~52, no.~11, pp.~104-110, Nov.~2014.

\bibitem{2013:Zhou}
X.~Zhou, R.~Zhang, and C.~K.~Ho, ``Wireless information and power transfer: architecture design and rate-energy tradeoff," \emph{IEEE Trans. Commun.}, vol.~61, no.~11, pp.~4754-4767, Nov. 2013.

\bibitem{2013:Zhang}
R.~Zhang and C.~K.~Ho, ``MIMO broadcasting for simultaneous wireless information and power transfer," \emph{IEEE Trans. Wireless Commun.}, vol.~12, no.~5, pp.~1989-2001, May 2013.

\bibitem{2015:Zeng}
Y.~Zeng and R.~Zhang, ``Optimized training design for wireless energy transfer," \emph{IEEE Trans. Commun.}, vol.~63, no.~2, pp.~536-550, Feb. 2015.

\bibitem{2015:Zeng1}
Y.~Zeng and R.~Zhang, ``Optimized training for net energy maximization in multi-antenna wireless energy transfer over frequency-selective channel," \emph{IEEE Trans. Commun.}, vol.~63, no.~6, pp.~2360-2373, Jun.~2015.

\bibitem{2014:Xu}
J.~Xu and R.~Zhang, ``Energy beamforming with one-bit feedback," \emph{IEEE Trans. Signal Process.}, vol.~62, no.~20, pp.~5370-5381, Oct. 2014.

\bibitem{2014:Huang1}
K.~Huang and V.~K.~N.~Lau, ``Enabling wireless power transfer in cellular networks: architecture, modeling and deployment," \emph{IEEE Trans. Wireless Commun.}, vol.~13, no.~2, pp.~902-912, Feb.~2014.

\bibitem{2015:Bi1}
S.~Bi and R.~Zhang, ``Placement optimization of energy and information access points in wireless powered communication networks," \emph{IEEE Trans. Wireless Commun.}, vol.~15, no.~3, pp.~2351-2364, Mar.~2016.


\bibitem{2014:Ju1}
H.~Ju and R.~Zhang, ``Throughput maximization in wireless powered communication networks," \emph{IEEE Trans. Wireless Commun.}, vol.~13, no.~1, Jan. 2014.

\bibitem{2014:Liu2}
L.~Liu, R.~Zhang, and K.~C.~Chua,  ``Multi-antenna wireless powered communication with energy beamforming," \emph{IEEE Trans. Commun.}, vol.~62, no.~12, pp.~4349-4361, Dec. 2014.

\bibitem{2015:Chen}
H.~Chen, Y.~Li, J.~L.~Rebelatto, B.~F.~Uchoa-Filho, and B.~Vucetic, ``Harvest-then-cooperate: wireless-powered cooperative communications," in \emph{IEEE Trans. Signal Process.}, vol.~63, no.~7, pp.~1700-1711, Feb. 2015.

\bibitem{2014:Zhou1}
X.~Zhou, R.~Zhang, and C.~K.~Ho, ``Wireless information and power transfer in multiuser OFDM systems," \emph{IEEE Trans. Wireless Commun.}, vol.~13, no.~4, pp.~2282-2294, Apr.~2014.

\bibitem{2013:Huang}
K.~Huang and E.~Larsson, ``Simultaneous information and power transfer for broadband wireless systems", \emph{IEEE Trans. Signal Processing}, vol.~61, no.~23, pp.~5972-5986, Dec. 2013.

\bibitem{2016:Zhou}
X.~Zhou, C.~K.~Ho, and R.~Zhang, ``Wireless power meets energy harvesting: a joint energy allocation approach in OFDM-based system," \emph{IEEE Trans. Wireless Commun.}, vol.~15, no.~5, pp.~3481-3491, May.~2016.

\bibitem{2013:Nintanavongsa}
P.~Nintanavongsa, M.~Y.~Naderi, and K.~R.~Chowdhury, ``Medium access control protocol design for sensors powered by wireless energy transfer," in \emph{Proc. IEEE INFOCOM}, pp.~150-154, Apr.~2013.

\bibitem{2013:Liu}
L.~Liu, R.~Zhang, and K.~C.~Chua, ``Wireless information transfer with opportunistic energy harvesting," \emph{IEEE Trans. Wireless Commun.}, vol.~12, no.~1, pp.~288-300, Jan.~2013.

\bibitem{2010:Grover}
P.~Grover and A.~ Sahai, ``Shannon meets Tesla: wireless information and power transfer," in \emph{Proc. IEEE ISIT}, pp.~2363-2367, Jun.~2010.

\bibitem{2014:Niyato}
D.~Niyato and P.~Wang, ``Competitive wireless energy transfer bidding: a game theoretic approach," in \emph{Proc. IEEE ICC}, Jun. 2014.

\bibitem{2001:Grimmett}
G.~Grimmett and D.~Stirzaker, \emph{Probability and random processes}, 3rd ed., Oxford University Press, New York, 2001.

\end{thebibliography}
\end{document}